# S-PATCH: MODIFICATION OF THE HERMITE PARAMETRIC PATCH

Václav SKALA and Vít ONDRAČKA
University of West Bohemia, Plzeň, Czech Republic

**ABSTRACT**: A new modification of the Hermite cubic rectangular patch is proposed - the S-Patch, which is based on the requirement that diagonal curves must be of degree 3 instead of degree 6 as it is in the case of the Hermite patch. Theoretical derivation of conditions is presented and some experimental results as well. The S-Patch is convenient for applications, where different tessellation of the u – v domain is needed, boundary and diagonal curves of different degrees are not acceptable.

**Keywords:** parametric surface, geometric modeling, computer graphics, spline, cubic surface.

………………………………………………………………………………………………………....

## 1. INTRODUCTION

Cubic parametric curves and surfaces are very often used for data interpolation or approximation. In the vast majority rectangular patches are used in engineering practice as they seem to be simple, easy to handle, compute and render (display). For rendering a rectangular patch is tessellated to triangles. In this paper we describe a new cubic patch modification, called Smart-patch (S-Patch). It is based on a Hermite cubic patch on which some additional requirements are applied. This modification is motivated by engineering applications, in general. It is expected that the proposed S-Patch can be widely applied in GIS systems and geography applications as well.

## 2. PROBLEM FORMULATION

Parametric cubic curves and surfaces are described in many publications [1-7]. There are many different formulas for cubic curves and patches, e.g. Hermite, Bezier, B-spline etc., but generally diagonal curves of a cubic rectangular patch are curves of degree 6. The proposed S-Patch, derived from the Hermite form, has diagonal curves of degree 3, i.e. curves for $v = u$ and $v = 1 - u$, while the original Hermite patch diagonal curves are of degree 6. Therefore the proposed S-Patch surface is "independent" of tessellation of regular $u - v$ domain. It means that if any tessellation is used, all curves, i.e. boundary and diagonal curves are of degree 3.

A cubic Hermite curve can be described in a matrix form as

$$x(t) = \boldsymbol{x}^T \boldsymbol{M}_H \boldsymbol{t} \qquad \boldsymbol{M}_H = \begin{bmatrix} 2 & -3 & 0 & 1 \\ -2 & 3 & 0 & 0 \\ 1 & -2 & 1 & 0 \\ 1 & -1 & 0 & 0 \end{bmatrix}$$

where: $\boldsymbol{x} = [x_1, x_2, x_3, x_4]^T$ is a vector of "control" values of a Hermite cubic curve, $x_3 = \frac{\partial x_1}{\partial t}$ and $x_4 = \frac{\partial x_2}{\partial t}$, $\boldsymbol{t} = [t^3, t^2, t, 1]^T$, $t \in \langle 0, 1 \rangle$ is a parameter of the curve and $\boldsymbol{M}_H$ is the Hermite matrix.

Hermite curve definition

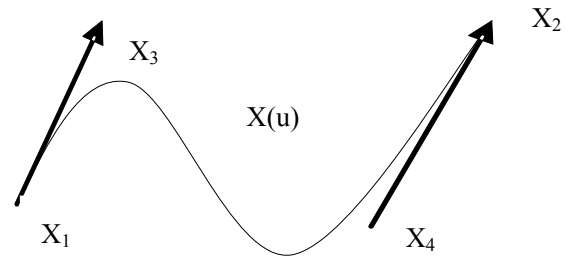

Figure 1: Hermite curve formulation

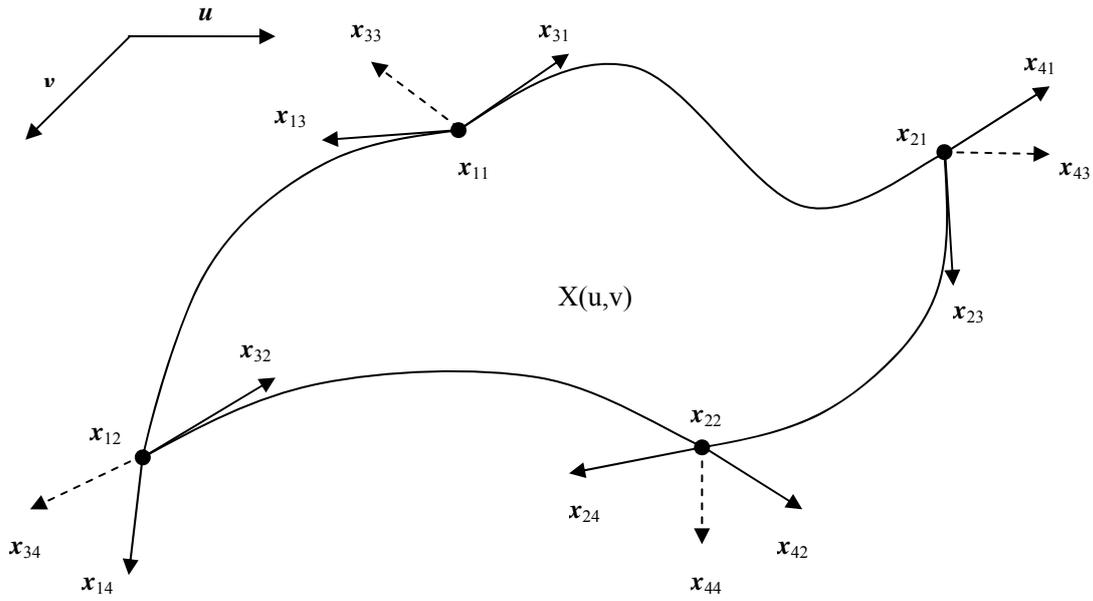

Figure 2: Hermite patch formulation

A cubic Hertmite patch (see Figure 2) is described in a matrix form for the $x$-coordinate as

$$x(u,v) = u^T M_H^T X M_H v$$

where: $X$ is a matrix of "control" values of the Hermite cubic patch

$$X = \begin{bmatrix} x_{11} & x_{12} & x_{13} & x_{14} \\ x_{21} & x_{22} & x_{23} & x_{24} \\ x_{31} & x_{32} & x_{33} & x_{34} \\ x_{41} & x_{42} & x_{43} & x_{44} \end{bmatrix} \quad \text{or}$$

$$X = \begin{bmatrix} x_{ij} & \frac{\partial x_{ij}}{\partial v} \\ \frac{\partial x_{ij}}{\partial u} & \frac{\partial^2 x_{ij}}{\partial u \partial v} \end{bmatrix} \text{ for i, j = 1,2}$$

$u$, resp. $v$ is a vector $u = [u^3, u^2, u, 1]^T$, resp. $v = [v^3, v^2, v, 1]^T$ and $u \in \langle 0,1 \rangle$, resp. $v \in \langle 0,1 \rangle$ is a parameters of the patch.

Similarly for $y$ and $z$ coordinates: $y(u,v) = u^T M_H^T Y M_H v$ and $z(u,v) = u^T M_H^T Z M_H v$. It means that a rectangular Hermite patch is given by a matrix 4 x 4 of control values for each coordinate, i.e. by 3 x 16 = 48 values in $E^3$.

From the definition of the Hermite patch it is clear, that boundary curves are cubic Hermite curves, i.e. curves of degree 3.

There are many applications, where a rectangular mesh is used in the $u - v$ domain. Sometimes $x$, resp. $y$ values are taken as $u$, resp. $v$ parameters and only $z$ value is interpolated/approximated as $y = f(x,y)$.

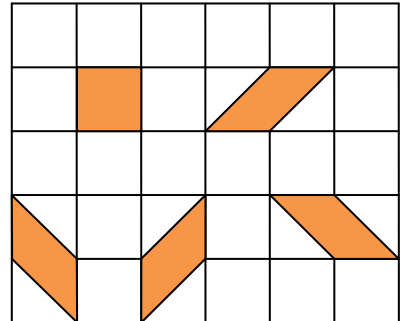

Figure 3: Different tessellation of u-v domain for the given corner points

There are many practical reasons why patches, i.e. the u-v domain, are tessellated to a triangular mesh, let us present just some of them:

1. A plane in $E^3$ is defined by three points, so the 4[th] point is not generally on the plane (due to computer limited precision it is nearly always valid even if the point theoretically lies on the plane).

2. The given $u - v$ rectangular domain mesh can be tessellated in different ways to a triangular mesh, in general, using different patterns (see Figure 3).

3. If a rectangular Hermite cubic patch is used, then the diagonal curves, i.e. $v = u$ and $v = 1 - u$, are of degree 6.

Especially last two points are very important as the final surface depends on tessellation and some curves might be of degree 6. This is not acceptable for some applications. There is a natural question: "Why some curves, i.e. when fixing $u = f(v)$, are of degree 3 and some are of degree 6?". If this feature is not controlled carefully it could lead to critical, sometimes even fatal, situations.

Understanding this, we exposed a specific restriction to the Hermite patch as curves for $v = u$ and $v = 1 - u$ must be of degree 3 as the patch boundary curves. This requirement has resulted into new modification of the Hermite cubic patch, called Smart-Patch (S-Patch), described below.

## 3. PROPOSED S-PATCH

Let us consider the Hermite patch on which we put some restrictions given by the requirement that diagonal curves, i.e. for $v = u$ and $v = 1 - u$, are to be of degree 3. The Hermite patch is given in the matrix form as

$$x(u,v) = \boldsymbol{u}^T \boldsymbol{M}_H^T \boldsymbol{X} \boldsymbol{M}_H \boldsymbol{v}$$

The restrictions for the proposed S-Patch are:
- $x(u,v)$ for $v = u$ is a curve of degree 3, it means that $x(u) = \boldsymbol{u}^T \boldsymbol{M}_H^T \boldsymbol{X} \boldsymbol{M}_H \boldsymbol{u}$ is a curve of degree 3.
  We can write $x(u) = \boldsymbol{u}^T \boldsymbol{R}_1 \boldsymbol{u}$, where $\boldsymbol{R}_1 = \boldsymbol{M}_H^T \boldsymbol{X} \boldsymbol{M}_H$

- $x(u,v)$ for $v = 1 - u$ is a curve of degree 3, it means that $x(u) = \boldsymbol{u}^T \boldsymbol{M}_H^T \boldsymbol{X} \boldsymbol{M}_H \boldsymbol{T} \boldsymbol{u}$ is a curve of degree 3, where:

$$\boldsymbol{v} = [(1-u)^3 \quad (1-u)^2 \quad 1-u \quad 1]^T$$

$$= \begin{bmatrix} -u^3 + 3u^2 - 3u + 1 \\ u^2 - 2u + 1 \\ -u + 1 \\ 1 \end{bmatrix} =$$

$$= \begin{bmatrix} -1 & 3 & -3 & 1 \\ 0 & 1 & -2 & 1 \\ 0 & 0 & -1 & 1 \\ 0 & 0 & 0 & 1 \end{bmatrix} \begin{bmatrix} u^3 \\ u^2 \\ u \\ 1 \end{bmatrix} = \boldsymbol{T}\, \boldsymbol{u}$$

We can write $x(u) = \boldsymbol{u}^T \boldsymbol{R}_2 \boldsymbol{u}$, where $\boldsymbol{R}_2 = \boldsymbol{M}_H^T \boldsymbol{X} \boldsymbol{M}_H \boldsymbol{T}$.

The Hermite diagonal curve is in both cases defined as

$$x(u) = \sum_{i,j=1}^{4} r_{ij}\, u^{4-i}\, u^{4-j} = r_{11} u^6$$
$$+ (r_{12} + r_{21})u^5$$
$$+ (r_{13} + r_{22} + r_{31})u^4$$
$$+ (r_{14} + r_{23} + r_{32} + r_{41})u^3$$
$$+ (r_{24} + r_{33} + r_{42})u^2$$
$$+ (r_{34} + r_{43})u + r_{44}$$

$$x(u) = \sum_{k=0}^{6} a_k\, u^k = \boldsymbol{a}^T \boldsymbol{u}$$

where:
$\boldsymbol{u} = [u^6, u^5, u^4, u^3, u^2, u, 1]^T$
$\boldsymbol{a} = [\, r_{11}\,,\, r_{12} + r_{21}\,,\, r_{13} + r_{22} + r_{31}\,,\, r_{14} + r_{23} + r_{32} + r_{41},\, r_{24} + r_{33} + r_{42},\, r_{34} + r_{43},\, r_{44}\,]^T$

We have 16 equations giving the relations between $x_{ij}$ and $r_{ij}$ for both cases, i.e. $\boldsymbol{R}_1$ and $\boldsymbol{R}_2$, that form a matrix relation, which expresses how the control values $x_{ij}$ form the coefficients in the $r_{ij}$

$$\boldsymbol{\rho} = \boldsymbol{\Omega}\, \boldsymbol{\xi}$$

where: $\boldsymbol{\rho}$ is a vector of coefficients of the matrix $\boldsymbol{R}$ and $\boldsymbol{\xi}$ is a vector of control points in the matrix $\boldsymbol{X}$.

For the case 1, i.e. $u = v$ we get

$$\xi = [\, x_{11},\ x_{12},\ x_{13},\ x_{14},\ x_{21},\ x_{22},\ x_{23},\ x_{24},\ x_{31},\ x_{32},\ x_{33},\ x_{34},\ x_{41},\ x_{42},\ x_{43},\ x_{44}\, ]^T$$

$$\Omega_1 = \begin{bmatrix}
4 & -4 & 2 & 2 & -4 & 4 & -2 & -2 & 2 & -2 & 1 & 1 & 2 & -2 & 1 & 1 \\
-6 & 6 & -4 & -2 & 6 & -6 & 4 & 2 & -3 & 3 & -2 & -1 & -3 & 3 & -2 & -1 \\
. & . & 2 & . & . & . & -2 & . & . & . & 1 & . & . & . & 1 & . \\
2 & . & . & . & -2 & . & . & . & 1 & . & . & . & 1 & . & . & . \\
-6 & 6 & -3 & -3 & 6 & -6 & 3 & 3 & -4 & 4 & -2 & -2 & -2 & 2 & -1 & -1 \\
9 & -9 & 6 & 3 & -9 & 9 & -6 & -3 & 6 & -6 & 4 & 2 & 3 & -3 & 2 & 1 \\
. & . & -3 & . & . & . & 3 & . & . & . & -2 & . & . & . & -1 & . \\
-3 & . & . & . & 3 & . & . & . & -2 & . & . & . & -1 & . & . & . \\
. & . & . & . & . & . & . & . & 2 & -2 & 1 & 1 & . & . & . & . \\
. & . & . & . & . & . & . & . & -3 & 3 & -2 & -1 & . & . & . & . \\
. & . & . & . & . & . & . & . & . & . & 1 & . & . & . & . & . \\
. & . & . & . & . & . & . & . & 1 & . & . & . & . & . & . & . \\
2 & -2 & 1 & 1 & . & . & . & . & . & . & . & . & . & . & . & . \\
-3 & 3 & -2 & -1 & . & . & . & . & . & . & . & . & . & . & . & . \\
. & . & 1 & . & . & . & . & . & . & . & . & . & . & . & . & . \\
1 & . & . & . & . & . & . & . & . & . & . & . & . & . & . & .
\end{bmatrix} \begin{bmatrix} r_{11} \\ r_{12} \\ r_{13} \\ r_{14} \\ r_{21} \\ r_{22} \\ r_{23} \\ r_{24} \\ r_{31} \\ r_{32} \\ r_{33} \\ r_{34} \\ r_{41} \\ r_{42} \\ r_{43} \\ r_{44} \end{bmatrix}$$

For the case 2, i.e. $v = 1 - u$ we get

$$\xi = [\, x_{11},\ x_{12},\ x_{13},\ x_{14},\ x_{21},\ x_{22},\ x_{23},\ x_{24},\ x_{31},\ x_{32},\ x_{33},\ x_{34},\ x_{41},\ x_{42},\ x_{43},\ x_{44}\, ]^T$$

$$\Omega_2 = \begin{bmatrix}
4 & -4 & 2 & 2 & -4 & 4 & -2 & -2 & 2 & -2 & 1 & 1 & 2 & -2 & 1 & 1 \\
-6 & 6 & -4 & -2 & 6 & -6 & 4 & 2 & -3 & 3 & -2 & -1 & -3 & 3 & -2 & -1 \\
. & . & . & -2 & . & . & . & -2 & . & . & . & -1 & . & . & . & -1 \\
. & 2 & . & . & . & -2 & . & . & . & 1 & . & . & . & 1 & . & . \\
6 & -6 & 3 & 3 & -6 & 6 & -3 & -3 & 4 & -4 & 2 & 2 & 2 & -2 & 1 & 1 \\
-9 & 9 & -3 & -6 & 9 & -9 & 3 & 6 & -6 & 6 & -2 & -4 & -3 & 3 & -1 & -2 \\
. & . & -3 & . & . & . & 3 & . & . & . & -2 & . & . & . & -1 & . \\
. & -3 & . & . & . & 3 & . & . & . & -2 & . & . & . & -1 & . & . \\
. & . & . & . & . & . & . & . & -2 & 2 & -1 & -1 & . & . & . & . \\
. & . & . & . & . & . & . & . & 3 & -3 & 1 & 2 & . & . & . & . \\
. & . & . & . & . & . & . & . & . & . & -1 & . & . & . & . & . \\
. & . & . & . & . & . & . & . & 1 & . & . & . & . & . & . & . \\
-2 & 2 & -1 & -1 & . & . & . & . & . & . & . & . & . & . & . & . \\
3 & -3 & 1 & 2 & . & . & . & . & . & . & . & . & . & . & . & . \\
. & . & . & -1 & . & . & . & . & . & . & . & . & . & . & . & . \\
. & 1 & . & . & . & . & . & . & . & . & . & . & . & . & . & .
\end{bmatrix} \begin{bmatrix} r_{11} \\ r_{12} \\ r_{13} \\ r_{14} \\ r_{21} \\ r_{22} \\ r_{23} \\ r_{24} \\ r_{31} \\ r_{32} \\ r_{33} \\ r_{34} \\ r_{41} \\ r_{42} \\ r_{43} \\ r_{44} \end{bmatrix}$$

We can write

$$a_i = \sum_{j=1}^{16} \lambda_{ij} \xi_j = \lambda_i^T \xi$$

for $i = 1, \dots, 6$.

As we require the diagonal curves to be of degree 3, we can write conditions for that as:

- The case 1:
  $$r_{11} = 0; \quad r_{12} + r_{21} = 0;$$
  $$r_{13} + r_{22} + r_{31} = 0$$
  using the matrix $R_1$

- The case 2:
  $$r_{11} = 0; \quad r_{12} + r_{21} = 0;$$
  $$r_{13} + r_{22} + r_{31} = 0$$
  using the matrix $R_2$

From those conditions we get a system of linear equations

$$\Lambda\, \xi = 0$$

where the first three rows of the matrix $\Lambda$ are taken for the case 1, i.e. related to the matrix $R_1$, and last three rows are taken for the case 2, i.e. related to the matrix $R_2$.

$$\xi = [\, x_{11},\ x_{12},\ x_{13},\ x_{14},\ x_{21},\ x_{22},\ x_{23},\ x_{24},\ x_{31},\ x_{32},\ x_{33},\ x_{34},\ x_{41},\ x_{42},\ x_{43},\ x_{44}\,]^T$$

$$\Lambda = \begin{bmatrix} 4 & -4 & 2 & 2 & -4 & 4 & -2 & -2 & 2 & -2 & 1 & 1 & 2 & -2 & 1 & 1 \\ -12 & 12 & -7 & -5 & 12 & -12 & 7 & 5 & -7 & 7 & -4 & -3 & -5 & 5 & -3 & -2 \\ 9 & -9 & 8 & 3 & -9 & 9 & -8 & -3 & 8 & -8 & 6 & 3 & 3 & -3 & 3 & 1 \\ -4 & 4 & -2 & -2 & 4 & -4 & 2 & 2 & -2 & 2 & -1 & -1 & -2 & 2 & -1 & -1 \\ 12 & -12 & 5 & 7 & -12 & 12 & -5 & -7 & 7 & -7 & 3 & 4 & 5 & -5 & 2 & 3 \\ -9 & 9 & -3 & -8 & 9 & -9 & 3 & 8 & -8 & 8 & -3 & -6 & -3 & 3 & -1 & -3 \end{bmatrix}$$

The rank of the matrix rank($\Lambda$) = 5, which means that we have to respect some restrictions generally imposed on the control points of the S-Patch. As the corner points are given by a user, the tangent and twist vectors are tied together with a relation. It can be seen that the vector $\xi$ is actually composed from values that are fixed (corner points are usually given) and by values, that can be considered as "free", but have to fulfill some additional condition(s). Let us explore this condition more in detail, now.

The equation $\Lambda\,\xi = 0$ can be rewritten as corner points are given as follows.

Let us define vectors $\xi_1$ and $\xi_2$, i.e. the corner points of the patch $\xi_1$ and tangent and twist vectors of the patch $\xi_2$, and and matrices $\Lambda_1$ and $\Lambda_2$ as

$$\xi_1 = [\, x_{11},\ x_{12},\ x_{21},\ x_{22}\,]^T$$

$$\Lambda_1 = \begin{bmatrix} 4 & -4 & -4 & 4 \\ -12 & 12 & 12 & -12 \\ 9 & -9 & -9 & 9 \\ -4 & 4 & 4 & -4 \\ 12 & -12 & -12 & 12 \\ -9 & 9 & 9 & -9 \end{bmatrix}$$

$$\xi_2 = [x_{13},\ x_{14},\ x_{23},\ x_{24},\ x_{31},\ x_{32},\ x_{33},\ x_{34},\ x_{41},\ x_{42},\ x_{43},\ x_{44}]^T$$

$$\Lambda_2 = \begin{bmatrix} 2 & 2 & -2 & -2 & 2 & -2 & 1 & 1 & 2 & -2 & 1 & 1 \\ -7 & -5 & 7 & 5 & -7 & 7 & -4 & -3 & -5 & 5 & -3 & -2 \\ 8 & 3 & -8 & -3 & 8 & -8 & 6 & 3 & 3 & -3 & 3 & 1 \\ -2 & -2 & 2 & 2 & -2 & 2 & -1 & -1 & -2 & 2 & -1 & -1 \\ 5 & 7 & -5 & -7 & 7 & -7 & 3 & 4 & 5 & -5 & 2 & 3 \\ -3 & -8 & 3 & 8 & -8 & 8 & -3 & -6 & -3 & 3 & -1 & -3 \end{bmatrix}$$

So we can write the equivalent to the equation $\Lambda\,\xi = 0$ as

$$[\Lambda_2 \quad \Lambda_1] \begin{bmatrix} \xi_2 \\ \xi_1 \end{bmatrix} = 0$$

As $\xi_1$ are given values (corner points of the patch) we can write

$$\Lambda_2 \xi_2 = -\Lambda_1 \xi_1$$

Rewriting and reducing the system of equations above, we get

$$\begin{bmatrix} x_{13} & x_{14} & x_{23} & x_{24} & x_{31} & x_{32} & x_{33} & x_{34} & x_{41} & x_{42} & x_{43} & x_{44} \\ 1 & . & -1 & . & . & . & . & . & 1 & -1 & 1 & . \\ . & 1 & . & -1 & . & . & . & . & 1 & -1 & . & 1 \\ . & . & . & . & 1 & -1 & . & . & -1 & 1 & -1 & -1 \\ . & . & . & . & . & . & 1 & . & . & . & . & 1 \\ . & . & . & . & . & . & . & 1 & . & . & 1 & . \end{bmatrix} \xi_2 =$$

$$\begin{bmatrix} x_{11} & x_{12} & x_{21} & x_{22} \\ -1 & 1 & 1 & -1 \\ -1 & 1 & 1 & -1 \\ -2 & 2 & 2 & -2 \\ 2 & -2 & -2 & 2 \\ 2 & -2 & -2 & 2 \end{bmatrix} \xi_1 \qquad (1)$$

From this equation (last two rows) we can see that the twist values of the patch must fulfill the following conditions:

$$x_{33} + x_{44} = 2\varphi$$
$$x_{34} + x_{43} = 2\varphi$$
$$\varphi = x_{11} - x_{12} - x_{21} + x_{22} \qquad (2)$$

where $\varphi$ is given by the corner points of the bicubic patch.

We can define two parameters $\alpha$ and $\beta$ (actually barycentric coordinates) as follows:

$$2\varphi\,\alpha = x_{44} \qquad 2\varphi\,(1-\alpha) = x_{33} \quad \text{and}$$
$$2\varphi\,\beta = x_{43} \qquad 2\varphi\,(1-\beta) = x_{34}$$

It means that the twist vectors are determined by $\alpha$ and $\beta$ values and by the corner points. Now we have the following equations to be solved (from the equation 1):

1st row $\qquad b + x_{43} = b + 2\beta\varphi = -\varphi$,

$$b = x_{13} - x_{23} + x_{41} - x_{42}$$

2nd row $\qquad a + x_{44} = a + 2\alpha\varphi = -\varphi$,

$$a = x_{14} - x_{24} + x_{41} - x_{42}$$

3rd row

$$c - x_{43} - x_{44} = c - 2(\alpha + \beta)\varphi = -2\varphi$$
$$c = x_{31} - x_{32} - x_{41} + x_{42}$$
$$\beta = -\frac{b+\varphi}{2\varphi} \qquad \alpha = -\frac{a+\varphi}{2\varphi}$$
$$-2\varphi = c - 2\varphi\alpha - 2\varphi\beta$$
$$= c + 2\varphi\frac{a+\varphi}{2\varphi} + 2\varphi\frac{b+\varphi}{2\varphi}$$
$$-2\varphi = c + a + \varphi + b + \varphi$$
$$a + b + c = -4\varphi$$

Expressing $\alpha$ and $\beta$ from the first two equations we get an equation (constraint) for the control values of the Hermite form that is actually the S-Patch as:

$$x_{31} - x_{32} + x_{41} - x_{42} + x_{14} - x_{24} - x_{23} + x_{13}$$
$$= -4\varphi$$

i.e.

$$x_{31} - x_{32} + x_{41} - x_{42} + x_{14} - x_{24} - x_{23} + x_{13} =$$
$$-4[x_{11} - x_{21} - x_{12} + x_{22}] \qquad (3)$$

This result should be read as follows:
- The S-Patch control points are determined parametrically. The control points (tangent vectors) of the border curves have to fulfill the condition above
- Twist vectors of the S-Patch are controlled by values $\alpha$ and $\beta$, that are determined from control points that form a boundary.

## 4. EXPERIMENTAL RESULTS

The experiments carried out proved that the proposed S-Patch has reasonable geometric properties.

To join S-Patches together a similar approach can be taken as for the standard Hermite rectangular patch, but there is just a small complication as the equation 1 has to be respected and kept valid. It has just influence to a curvature of a surface of the neighbors.

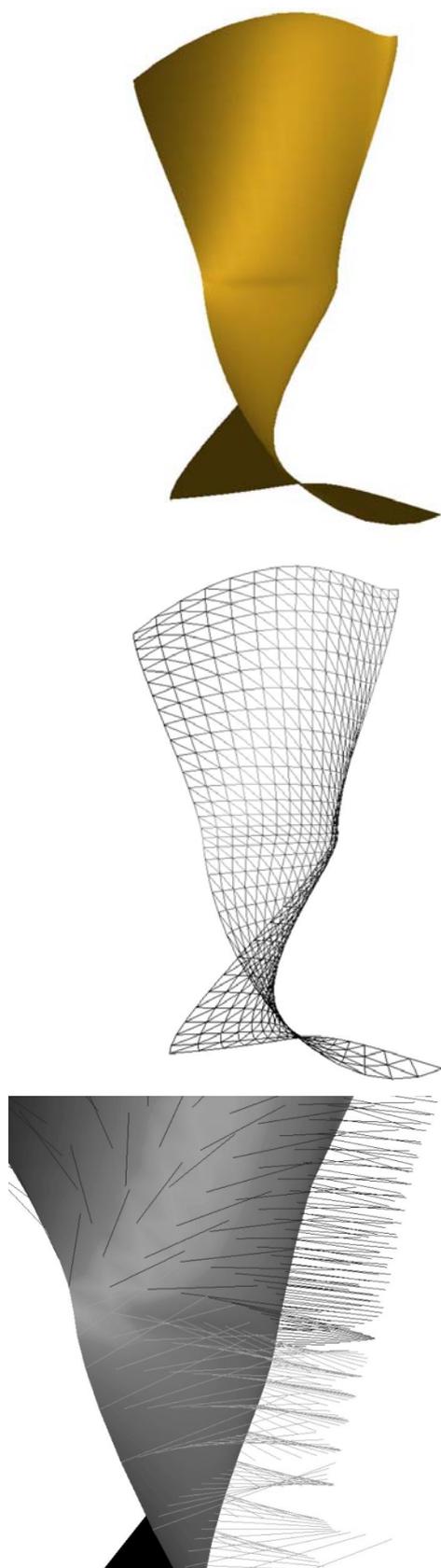

Figure 4: Two joined patches rendered, mesh and normal vectors

The experiments made proved that it is possible to join S-Patches smoothly. To prove additional basic properties of the proposed S-Patch we used ½ of a cube.

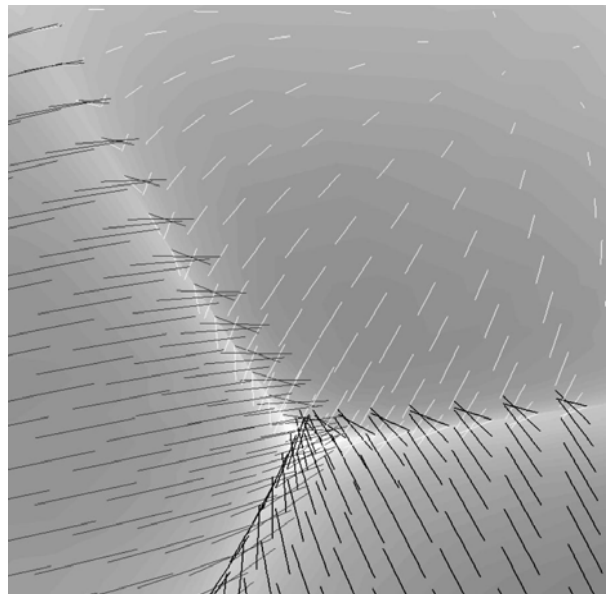

Figure 5: Joined patches of ½ cube

Nevertheless there was a severe problem detected, when the vertex of a mesh is shared by three patches. In some cases it was difficult to keep $C^1$ continuity (see Figure 5).

## 5. CONCLUSION

We have described and derived a new modification of the Hermite cubic patch. The main advantages of the proposed S-Patch are:

- Twist vectors are determined by the equation 1 due to the restriction put on diagonal curves as we require degree 3.
- Both diagonal curves are cubic curves, i.e. curves of degree 3.
- Different tessellations of $u-v$ domain and conversion to triangles do not change the degree of border and diagonal curves.
- Curves (boundary and diagonal) are of degree 3, less operations are needed as the computed polynomial is of degree 3.
- The given $u-v$ domain can be tessellated in different ways to four sided mesh and to triangular meshes for rendering using different tessellations.

It should be noted that one additional

condition, i.e. equation 1 has to be kept valid, that complicates the implementation a little bit, but on the other hand the presented advantages of the S-Patch seem to be obvious.

The future work is to be especially devoted to:

- Exploration how the S-Patch formulation should be modified for cases when a vertex of the mesh is shared by 3 or 5 patches.
- Analysis of the S-Patch properties (smoothness, curvature etc.)
- Derivation of S-Patch conditions for the Bezier and other patch types.
- Methods for users interface implementation including user interaction and manipulation in surface design.

There is a belief, that it will be possible to derive additional general properties of the proposed S-Patch as well.

**ACKNOWLEDGMENTS**

The authors would like to express thanks to colleagues at the Center of Computer Graphics and Visualization for challenging this work, for many comments and hints they have made. Also comments by anonymous reviewers were constructive and helped to improve this contribution a lot.

The project was supported by the Ministry of Education of the Czech Republic, project VIRTUAL, No.2C06002.

**ABOUT THE AUTHORS**

1. Václav SKALA, is a professor at the University of West Bohemia, the head of the Center of Computer Graphics and Visualization (http://Graphics.zcu.cz) and is the co-chair of the WSCG (http://wscg.zcu.cz) and GraVisMa (http://GraVisMa.zcu.cz) conferences. His e-mail and postal address is as follows: skala@kiv.zcu.cz Department of Computer Science and Engineering Studies, Faculty of Applied Sciences, University of West Bohemia, CZ 306 14 Plzen (Pilsen), Czech Republic.

2. Vít ONDRAČKA is a research fellow at the Center of Computer Graphics and Visualization with the Department of Computer Science and Engineering Studies, Faculty of Applied Sciences, University of West Bohemia, CZ 306 14 Plzen (Pilsen), Czech Republic.